\documentstyle[amstex,prl,aps,epsf,multicol,epsfig]{revtex}

\begin{document}
\title{Decoherence Correction by the Zeno Effect and Non-Holonomic Control
}
\author{E.~Brion,$^{1,*}$ G.~Harel,$^{2,3}$ N.~Kebaili,$^{1}$
 V.~M.~Akulin,$^{1,\dagger}$ and I.~Dumer$^{4,\ddagger}$} 
\address{$^{1}$Laboratoire Aim\'{e} Cotton, CNRS II,
B\^atiment 505, 91405 Orsay Cedex, France\\
$^{2}$Spinoza Institute, Utrecht University, Leuvenlaan 4,
3508 TD Utrecht, The Netherlands\\
$^{3}$Department of Computing, University of Bradford, Bradford BD7 1DP, United Kingdom\\
$^{4}$College of Engineering, University of California, Riverside, CA 92521, USA}
\date{\today}
\maketitle
\widetext
\begin{abstract}
We show that multidimensional Zeno effect combined with non-holonomic
control allows to efficiently protect quantum systems from decoherence by
a method similar to classical coding.
Contrary to the conventional approach, our method is applicable to arbitrary error-inducing Hamiltonians and general quantum systems. 
We also propose algorithms of finding encoding that approaches 
the Hamming upper bound along with methods of practical realizations of the encodings. Two new codes protecting 2 information qubits out of 7 and 4 information qubits out of 9 against a single error with arbitrarily small probability of failure are constructed as an example. 
\end{abstract}
\draft
\begin{multicols}{2}
\narrowtext
Decay and decoherence are the two processes that occur in open quantum
systems. Two different relaxation times $T_{1}$ and $T_{2}$ characterize
these processes in two-level systems. Though both processes represent
irreversible relaxation, they are of a rather different nature. The number
of occupied states given by the rank of the density matrix - an analog of
the phase volume in the classical case - may change in an elementary act of
relaxation, and usually does it towards diminishing. By contrast, this
number does not change in an elementary act of decoherence, but yet
increases after being averaged over subsequent elementary acts.

In this Letter, we show how one can completely suppress decoherence in a
universal way. We employ basic ideas of classical error-correcting codes and
their powerful extensions \cite{Calderbank:96} that make use of extraspecial and
Clifford groups to build quantum codes from their classical counterparts.
However, our approach considerably deviates from this method, since it
relies on the full group of unitary transformations in the entire Hilbert
space. We also employ the idea known as the Zeno effect\cite{zeno} that
states that periodic measurements allow to keep a quantum system in the
initial state for as long as needed, provided that the period $T$ is short
relative to the relaxation time $T_{2}$. However, contrary to the
conventional Zeno effect, we do not intervene into the entire quantum
system. Instead, we only affect the auxiliary part, the ancilla, to protect the
quantum state of the main part. As a general tool for
protecting any system from decoherence, we suggest the non-holonomic control
\cite{Non-holonomic}, which allows one to perform any predetermined
transformation including encoding in the Hilbert space of a quantum
system.

An important example emerges when a classical binary code is compared with
its conventional quantum counterpart \cite{Calderbank:96,Nielsen}. Both
codes add the auxiliary part to increase minimal discrepancy between possible
states. However, classical codes use a finite alphabet mapping $k$
information bits onto a single code vector of a bigger length $n$. By
contrast, the state of $k$ qubits is a continuous complex-valued vector
in the entire $2^{k}$-dimensional Hilbert space ${\cal H}_{k}.$ 
When the original system is expanded by $n-k$ ancilla qubits in some state $%
\left| \widetilde{\alpha }\right\rangle $, the extended state vectors form
a subspace ${\cal H}_{k,\widetilde{\alpha }}=\left| 
\widetilde{\alpha }\right\rangle \left\langle \widetilde{\alpha }\right| 
{\cal H}_{n}$ in the bigger space ${\cal H}_{n}$. Quantum encoding $\widehat{%
C}$ is a unitary transformation that maps the subspace ${\cal H}_{k,%
\widetilde{\alpha }}$ onto a $2^{k}$ dimensional code subspace ${\cal C}=%
\widehat{C}{\cal H}_{k,\widetilde{\alpha }}$ in ${\cal H}_{n}$. Note that $%
{\cal C}$ spans over all columns of the rectangular matrix $\widehat{C}%
\left| \widetilde{\alpha }\right\rangle $, and has zero projections \ on all 
$2^{n}-2^{k}$ vectors that form the columns of \ any other matrix $\widehat{C%
}\left| \alpha\right\rangle$
with $|\alpha\rangle \perp |\widetilde{\alpha }\rangle$ in the ancilla.
\marginpar{
?}

There is also a difference in the decoding process. A corrupted codeword can
be fully retrieved and corrected by a classical code. By contrast, in
quantum mechanics, retrieval of the entire state vector by a measurement
destroys the data. A natural way of quantum correction is therefore to find
an encoding $\widehat{C}$ and decoding $\widehat{C}^{-1}$ that can move all
possible errors to the ancilla. In this case, resetting the ancilla to the
state $\left| \widetilde{\alpha }\right\rangle $ gives a projection of the
entire state vector exactly to its original position in the subspace ${\cal H}_{k,\widetilde{\alpha }}$. To this end, given any set of errors $\left\{ \widehat{E}_{m}\right\} $,
we should find a code ${\cal C=}\{\left| v\right\rangle \}$ such that all
possible error vectors $\widehat{E}_{l}\left| v\right\rangle $ are
orthogonal to ${\cal C}$ and to each other\cite{Knill} according to the
condition 
\begin{equation}
\left\langle v^{\prime }\right| \widehat{E}_{s}\widehat{E}_{l}\left|
v\right\rangle =0;\ \ \forall \left| v\right\rangle ,\left| v^{\prime
}\right\rangle \in {\cal C};\ \ \forall \widehat{E}_{s},\widehat{E}_{l}\in
\left\{ \widehat{E}_{m}\right\} .  \label{EQ1a}
\end{equation}
Formally, this requirement coincides with the definition of a code 
${\cal C}$ (see \cite{Nielsen} p.\ 436) written for zero trace matrices 
$\widehat{E}_{m}$\cite{condition}, although the errors we consider are not
restricted to the
extraspecial group in contrast to the original formulation. It can be seen
that Eq.(\ref{EQ1a}) also corresponds to the classical Gilbert-Varshamov 
\cite{Varshamov} bound.

Below we describe a recipe for constructing codes that are ``nearly'' orthogonal to error vectors with
any predetermined accuracy, so that the corrupted state vector falls
arbitrarily close to the original code vector, after being projected on the
code subspace. It allows one to replace the rather strict condition Eq.(\ref
{EQ1a}) by a weaker condition,
\begin{equation}
\left\langle v^{\prime }\right| \widehat{E}_{m}\left| v\right\rangle
=0;\ \ \forall \left|v\right\rangle,\left| v^{\prime }\right\rangle\in{\cal C};
\ \ \forall \widehat{E}_{m}\in \left\{ \widehat{E}_{m}\right\} .  \label{EQ2a}
\end{equation}
Note that the latter can be thought of as an analog of the classical Hamming
bound\cite{Varshamov}. We emphasize that our approach extends beyond the
quantum systems composed of identical two-level particles and is valid for
any quantum system separable into a main part and an ancilla. In fact, this
approach only depends on the total number $M$ of the error-inducing
Hamiltonians $\left\{\widehat{E}_{m}\right\}$
regardless of their specific matrix structure.

To design codes that meet the conditions of Eq.(\ref{EQ2a}), we will use the
Zeno effect. The regular Zeno effect allows one to restore the initial state
vector up to the second-order terms in $T/T_{2}$, which are very small
provided that $T\ll T_{2}$. The essence of this phenomenon is based on the
heuristic that within a short time interval, a unitary quantum evolution moves
any state vector in a direction orthogonal to the vector itself. In our
version of the Zeno effect, the subsequent projection is done to the $2^{k}$%
-dimensional subspace ${\cal H}_{k,\widetilde{\alpha }}$ instead of the
state vector. Therefore, our primary goal is to find a code ${\cal C}$ such that
for short time intervals, any error Hamiltonian in $\left\{ \widehat{E}%
_{m}\right\} $ moves any code vector in a direction perpendicular to ${\cal C%
}$.

To implement the Zeno effect, consider the state vector $\left|
S\right\rangle =\left| s\right\rangle \otimes \left| \widetilde{\alpha }%
\right\rangle $ of the compound system formed by the information part in the
state $\left| s\right\rangle $ and the ancilla in the state $\left| 
\widetilde{\alpha }\right\rangle $. The vector $\left| S\right\rangle $
undergoes an uncontrolled unitary evolution 
\begin{equation}
\widehat{U}_{E}=\prod_{m=1}^{M}e^{-i\widehat{E}_{m}\int
\!\!f_{m}(t)dt}\simeq \widehat{I}-i\sum_{m=1}^{M}\widehat{E}_{m}\int
\!\!f_{m}(t)dt  \label{EQ2}
\end{equation}
where the time ordering of the product is implicit. To illustrate the idea
of the approach, we simply treat the errors $\widehat{E}_{m}$ as
Hamiltonians of interactions with external random fields $f_{m}(t)$ that
produce an uncontrolled evolution of the system. We also assume that over
the Zeno period $T$ the different fields have actions $\int_{t}^{t+T}\!%
\!f_{m}(x)dx\left| \widehat{E}_{m}\right| \ll 1$ that are so small that only
the identity operator $\widehat{I}$ and the first order terms $\sim \int
\!\!f_{m}(t)dt$ are important in the Taylor series, whereas the higher orders
can be ignored.

Let $\widehat{\rho }_{sc}=\left| s\right\rangle \otimes \left\langle
s\right| $ be the density matrix of the main part before the action
of errors. Then $\widehat{\rho }=\left| S\right\rangle \otimes \left\langle
S\right| =\left| \widetilde{\alpha }\right\rangle \widehat{\rho }%
_{sc}\left\langle \widetilde{\alpha }\right| $ is the density matrix of the
entire system. The variation $\delta \widehat{\rho }_{sc}=\left\langle 
\widetilde{\alpha }\right| \delta \widehat{\rho }\left| \widetilde{\alpha }%
\right\rangle $ of the density matrix of the main system after the
perturbation Eq.(\ref{EQ2}) and resetting of the ancilla is given by the
commutator 
\begin{equation}  \label{EQ9}
\delta \widehat{\rho }_{sc}=-i\left[ \sum_{m=1}^{M}\int
\!\!f_{m}(t)dt\left\langle \widetilde{\alpha }\right| \widehat{C}^{-1}%
\widehat{E}_{m}\widehat{C}\left| \widetilde{\alpha }\right\rangle ,\widehat{%
\rho }_{sc}\right] .
\end{equation}
Therefore $\widehat{\rho }_{sc}$ satisfies the master equation 
\begin{equation}  \label{EQ10}
i\frac{d\widehat{\rho }_{sc}}{dt}=\left[ \widehat{h}_{e},\widehat{\rho }_{sc}%
\right];\ \ \widehat{h}_{e}=\sum_{m=1}^{M}f_{m}\left\langle \widetilde{\alpha }%
\right| \widehat{C}^{-1}\widehat{E}_{m}\widehat{C}\left| \widetilde{\alpha }%
\right\rangle
\end{equation}
with effective Hamiltonian $\widehat{h}_{e}$. Then requirement Eq.(\ref{EQ2a}%
) implies that $\widehat{h}_{e}=0$ and $\widehat{\rho }_{sc}=$const, whereas
the relation $\left| v\right\rangle =\widehat{C}\left| s\right\rangle
\otimes \left| \widetilde{\alpha }\right\rangle $ holds for the code vectors
and the states of the system.%
\marginpar{
same?}

Our next step is to design codes that satisfy Eq.(\ref{EQ2a}). Let $N=2^{k}$
and $A=2^{n-k}$ be the dimensions of the Hilbert spaces formed by the
main part (information subsystem) and ancilla, respectively. Then Eq.~(\ref{EQ2a}) is a
set of $M\times N^{2}$ equations taken over $A\times N^{2}$ matrix elements $%
\left\langle \alpha \right| \otimes \left\langle s^{\prime }\right| \widehat{%
C}\left| s\right\rangle \otimes \left| \widetilde{\alpha }\right\rangle $ of
the encoding operator that define the rectangular $N\times NA$ matrix $%
\widehat{C}\left| \widetilde{\alpha }\right\rangle $. To solve this system,
we wish to limit the number of equations by the number of \ free variables.
This gives the Hamming upper bound in the form $M\leq A.$ To specify Eq.~(%
\ref{EQ2a}) further, let the states $\left| s\right\rangle \otimes \left| 
\widetilde{\alpha }\right\rangle $ correspond to the first $N$ positions of
the state vector, which implies that each of the $M$ matrices $\widehat{C}%
^{-1}\widehat{E}_{m}\widehat{C}$ has all-zero upper left $N\times N$ corner.
Standard methods of linear algebra do not give a recipe of finding a linear
transformation $\widehat{C}$ that simultaneously sets to zero the corners of 
$M$ different matrices. The first result of the paper is an iterative
algorithm which does so, and hence finds a code ${\cal C}$ that satisfies the
conditions of Eq.(\ref{EQ2a}) given $M$ arbitrary error Hamiltonians
$\widehat{E}_{m}$.

To illustrate the method, we first find a single vector $\left|
x\right\rangle $ of length $N\times A$ that is orthogonal to all $M$ error
vectors $\widehat{E}_{m}\left| x\right\rangle $. In other words, $\left|
x\right\rangle $ should satisfy $M$ quadratic equations 
\begin{equation}
\left\langle x\right| \widehat{E}_{m}\left| x\right\rangle =0.  \label{EQ22}
\end{equation}
Our input is a randomly selected vector $\left| x\right\rangle $ of unit
length. If, by chance, $\left| x\right\rangle $ is orthogonal to all vectors 
$\widehat{E}_{m}\left| x\right\rangle $, the problem is solved. Otherwise,
there exists a linear combination $\sum_{m=1}^{M}\gamma _{m}\widehat{E}%
_{m}\left| x\right\rangle $ that minimizes the length of the vector $%
\sum_{m=1}^{M}\gamma _{m}\widehat{E}_{m}\left| x\right\rangle +\left|
x\right\rangle $. To find it, we employ the standard variational principle
and solve the corresponding set of linear equations for $\gamma _{m}$. In
our further iterations, we set $\left| x\right\rangle \rightarrow
\sum_{m=1}^{M}\frac{1}{2}\gamma _{m}\widehat{E}_{m}\left| x\right\rangle
+\left| x\right\rangle $, then normalize the new vector to unity, and repeat
the procedure until it converges. \ The same algorithm serves to find the
entire encoding matrix $\widehat{C}$. \ For this purpose, one should simply
consider a supervector $x=\overrightarrow{\widehat{C}\left| \widetilde{%
\alpha }\right\rangle }$ of length $N^{2}A$ by appending the columns of $%
\widehat{C}\left| \widetilde{\alpha }\right\rangle $. Then the orthogonality
conditions $\left\langle \nu _{l}\right. \left| \nu _{s}\right\rangle =0$
taken for the different columns\ $s,l,$ $s\neq l,$ and the conditions $%
\left\langle \nu _{l}\right| \widehat{E}_{m}\left| \nu _{s}\right\rangle =0$
give independent equations for this supervector.

This procedure has shown a rapid rate of convergence for all the examples
considered, including two new single-error correcting codes with
$(n,k)$ parameters $%
\left( 7,2\right) $ and $\left( 9,4\right) $\cite{brion-WEB}$.$ Note that
these codes are designed in the Hilbert space and perform single error
correction with any predetermined accuracy that depends only on $T$.
However, neither of the two\ codes can exist for conventional quantum
techniques that satisfy  Eq.(\ref{EQ1a}). More generally, our conjecture is
that the above procedure converges if the length of a supervector $x$ is no
less than the number of quadratic equations Eq.(\ref{EQ22}), or,
equivalently, if the Hamming bound $M\leq A$ holds. For this conjecture,
note also that all matrices $\widehat{E}_{m}$ in Eq.(\ref{EQ22}) have zero
traces.

Once a code ${\cal C}$ is found, a question arises as to how the system can be
moved into this subspace. Thus, we also need to design an efficient encoding 
$\widehat{C}.$ For the codes built on the Clifford group, this can be done
by applying a polynomial number of standard operations like Hadamard gates.
It is not the case for an arbitrary quantum system, since for generic error
sets $\left\{\widehat{E}_{m}\right\}$,
$\widehat{C}$ is also a generic unitary matrix, and
hence an alternative method of coding is required. One can achieve this
feasibility with the help of non-holonomic control, by applying a sequence
of two different natural Hamiltonians $\widehat{H}_{1}$ and $\widehat{H}_{2}$
that satisfy certain conditions\cite{Non-holonomic}. The second result of
the paper is an algorithm that allows one to determine a set of
non-holonomic control timings $t_{1},t_{2},t_{3},\cdots ,t_{M^{\prime }}$ of
a reasonably short length $M^{\prime }$ resulting in the encoding
transformation 
\begin{equation}  \label{EQ0}
\widehat{C}=e^{-it_{M^{\prime }}\widehat{H}_{2}}\cdots e^{-it_{3}\widehat{H}%
_{1}}e^{-it_{2}\widehat{H}_{2}}e^{-it_{1}\widehat{H}_{1}}.
\end{equation}
The number $M^{\prime }$ equals or slightly exceeds the total number $MN^{2}$
of relevant error matrix elements.

The key idea of the method is to incorporate the search for the
non-holonomic control parameters $t_{l}$ into the iterative algorithm for
the constants $\gamma _{m}$. 
\marginpar{
?} Being written in the form of an $N\times N$
matrix $\widehat{\gamma }_{m},$ these
constants relate the supervector $\overrightarrow{\widehat{C}\left| 
\widetilde{\alpha }\right\rangle }$ to its change $\sum_{m=1}^{M}%
\overrightarrow{\widehat{E}_{m}\widehat{C}\left| \widetilde{\alpha }%
\right\rangle \widehat{\gamma }_{m}}$, thus allowing us to find the
first-order variations $\!\widehat{\delta }_{m}=\left\langle \widetilde{%
\alpha }\right| \widehat{C}^{-1}\widehat{E}_{m}\widehat{C}\left| \widetilde{%
\alpha }\right\rangle \widehat{\gamma }_{m}$ $+\widehat{\gamma }%
_{m}^{\dagger }\left\langle \widetilde{\alpha }\right| \widehat{C}^{-1}%
\widehat{E}_{m}^{\dagger }\!\widehat{C}\left| \widetilde{\alpha }%
\right\rangle $ of the error matrices projected on the code space. We
therefore choose the variations $\delta t_{l}$ of the control timings such
that the corresponding increments of the error matrices are equal to $\widehat{%
\delta }_{m}$ for each $m$, that is, \smallskip\ 
\begin{align}
& \left\langle \widetilde{\alpha }\right| \widehat{C}^{-1}\widehat{E}_{m}%
\widehat{C}\left| \widetilde{\alpha }\right\rangle \widehat{\gamma }_{m}+%
\widehat{\gamma }_{m}^{\dagger }\left\langle \widetilde{\alpha }\right| 
\widehat{C}^{-1}\widehat{E}_{m}^{\dagger }\!\widehat{C}\left| \widetilde{%
\alpha }\right\rangle  \nonumber \\
& =\beta \sum_{l=1}^{M^{\prime }}\left( \left\langle \widetilde{\alpha }%
\right| \widehat{C}^{-1}\widehat{E}_{m}\frac{\partial \widehat{C}}{\partial
t_{l}}\left| \widetilde{\alpha }\right\rangle +\left\langle \widetilde{%
\alpha }\right| \frac{\partial \widehat{C}^{-1}}{\partial t_{l}}\widehat{E}%
_{m}^{\dagger }\!\widehat{C}\left| \widetilde{\alpha }\right\rangle \right)
\delta t_{l}.\smallskip  \label{EQ16}
\end{align}
Then we continue the iterations, while varying the big parameter $\beta $ to
gain maximum convergency. The number $M^{\prime }$ of the parameters $%
t_{l}$ can be chosen slightly above the minimum number $MN^{2}$ in order to
avoid a premature stop if the $MN^{2}\times MN^{2}$ supermatrix in the
parentheses of \ Eq.(\ref{EQ16}) becomes singular for some intermediate
approximation $\widehat{C}$ of the coding operator.

This error-correction method becomes asymptotically exact for $T\rightarrow
0 $. However promising for small quantum systems, it has complexity that is
exponent in the number of particles $n$. Therefore, we also propose an
alternative strategy applicable to large systems, which is the third result
of the paper. If the computational time required to find the $MN^{2}\times
MN^{2}$ matrix becomes prohibitive, thus hindering any way to exactly meet
the conditions of Eq.(\ref{EQ2a}), one can minimize the rate of error
accumulation in the density matrix, by minimizing the effective Hamiltonians
in Eq.(\ref{EQ10}). For the systems with binary interaction an exponential
reduction of the error-accumulation rate can be achieved if the projection
rate $1/T$ grows faster than $n^{2}$. Moreover, the code rate $R=k/n$ tends
to unity as $1-o(n^{2}T).$

In particular, one can reduce all effective Hamiltonians $\widehat{h}%
_{m}=f_{m}\left\langle \widetilde{\alpha }\right| \widehat{C}^{-1}\widehat{E}%
_{m}\widehat{C}\left| \widetilde{\alpha }\right\rangle $ by applying a
random encoding $\widehat{C}$, which allows to inhibit the rate of
decoherence by the order of ancilla dimension $A=2^{n-k}$. \ Indeed, for
individual qubits, a typical single-particle or binary interaction $\widehat{%
E}_{m}$ is given by a sparse matrix that strongly couples each state to 
sufficiently few other states. By contrast, most matrix elements are zeros.
A generic unitary encoding $\widehat{C}$ smears out this coupling over all $%
2^{n}$ states of the entire system, thus making a typical matrix element of $%
\widehat{C}^{-1}\widehat{E}_{m}\widehat{C}$ \;$2^{n}$ times smaller than the
typical nonzero matrix element of $\widehat{E}_{m}$. \ Since only $2^{k}$ of
these states contribute to the projected interactions $\widehat{h}_{m},$ the
overall gain is equal to $2^{n-k}$. \ From a more general viewpoint, this
random encoding rests on the fact that the decoherence rate is a polynomial
in the number $n$ of particles, whereas the dimension of the entire Hilbert
space is an exponent in $n$. It allows to store quantum information in
strongly entangled states of many-body systems that are only weakly affected
by physically realistic perturbations.

This random encoding also allows to specify a relation between the dimension 
$N=2^{k}$ of the information system and the maximum number $t$ of possible
errors caused by binary interactions. Given $K$ single-particle quantum
states, there exist $M=\sum_{l=1}^t\left(_{l}^{n}\right)\left(K^2-1\right)^{l}$
different possible error matrices (here $K=2$ for qubits). \ Since the
energy of binary interaction scales as $n^{2}$, each error matrix yields a
decoherence rate of $W=n^{2}/T_{2}$.  This rate is reduced $A=2^{n-k}$ 
times after the matrix elements are smeared out over the entire system.
Thus, given $n\rightarrow \infty $ identical particles, one can correct $t$
errors with high fidelity if the total decoherence rate $W2^{k-n}M$
vanishes. This condition $W2^{k-n}M\rightarrow 0$ again gives the Hamming
bound \cite{Varshamov} 
\begin{equation}
1-k/n-\log _{2}M^{1/n}\geq 0.  \label{EQ91}
\end{equation}
Hence, to asymptotically meet this classical bound, we can randomly choose
an $(n,k)$-code thus leaving the error patterns of weight $t$ or less
uncorrected with a vanishing probability.

Finally, note that the sequential application of two different Hamiltonians
suggested by Eq.(\ref{EQ0}) results in a completely generic transformation,
provided that: (i) the number $M^{\prime }$ of the control timings $t_{l}$
is of the order of $\log _{2}\left( N\times A\right) $; (ii) the operators $%
\widehat{H}_{1}$ and $\widehat{H}_{2}$ satisfy the conditions of
non-holonomic control\cite{Non-holonomic}; and (iii) the time intervals $%
t_{l}$ are long enough to ensure big acquired actions $||\widehat{H}%
_{1,2}\,t_{l}||\gg \hbar $. Therefore, given a system, the main problem is
to find two controllable interactions that satisfy the conditions needed for
non-holonomic control, and allow to perform the decoding transformation $%
\widehat{C}^{-1}$ at the same level of complexity.

We present such a pair $\widehat{H}_{1}$, $\widehat{H}_{2}$ for the example
of the molecule proposed in \cite{Vandersypen:01} as a toy quantum computer
of 7-qubits, for which  the applicability of both the exact and the
approximative approaches was explicitly verified by  determining  the
control timings and the reduction of effective coupling, respectively. The
first operator is the nuclear magneto-dipole interaction 
\begin{equation}
\widehat{H}_{1}=\sum_{j}B_{x}\mu _{j}\widehat{\sigma }_{x}^{(j)}  \label{EQ7}
\end{equation}
of the molecule in the ground ro-vibronic state $\left| g\right\rangle $
with a static external magnetic field $B_{x}$ oriented along the $x$-axis.
The second operator is the Raman coupling 
\begin{equation}
\widehat{H}_{2}=\sum_{i<j}\sum_{a,b=x,y,z} \delta\omega^{-1}\mu _{i,j}^{(2)}B^{R}_{a}B^{R}_{b}\,%
\widehat{\sigma}_{a}^{(i)}\!\otimes\widehat{\sigma}_{b}^{(j)}\label{EQ8}
\end{equation}
in a field $\overrightarrow{B^{R\ }}$ oscillating at a high-frequency $%
\omega $ detuned by $\delta \omega =\omega _{tr}-\omega $ from the frequency 
$\omega _{tr}$ of transition to an exited ro-vibronic state $\left|
e\right\rangle $. Here $\mu _{j}=\left\langle g\right| \mu _{j}(r)\left|
g\right\rangle $ is the average value of the coordinate dependent
gyromagnetic ratio $\mu _{j}(r)$ of the $j$-th nucleus in the ground state,
and $\mu _{i,j}^{(2)}=\left\langle g\right| \mu _{i}(r)\left| e\right\rangle
\left\langle e\right| \mu _{j}(r)\left| g\right\rangle $ represents the
matrix element of the Raman transition. By changing the direction of $B_{x}$
from positive to negative and reversing the detuning $\delta \omega $, one
can vary the signs of the Hamiltonians, which allows one to  easily
construct the inverse transformation $\widehat{C}^{-1}$ just by inverting
the order of the sequence $t_{M^{\prime }},\dots ,t_{1}$ in Eq.(\ref{EQ0})
along with the change of the signs.

We do not present here\cite{brion-WEB} a cumbersome set of $M^{\prime }\sim
2^{5+2\times 2}$ non-holonomic control timings found explicitly with the
help of the approximative algorithm for the above single error-correcting
code that protects two out of seven qubits in the toy computer. Nor do we
give the explicit matrix elements of the effective error Hamiltonians Eq.(%
\ref{EQ10}) obtained as a result of ten sequential applications of $\widehat{%
H}_{1}$ and $\widehat{H}_{2}$ that reduce the decoherence rate of the two
selected qubits. We only note that our numeric results are completely
consistent with the given estimates, whereas the absolute values of the
matrix elements obey a Maxwellian-like statistics.

Summarizing the main results of the paper, we consider open quantum systems
that consist of two parts - the main part and an ancilla. The latter is being
periodically reset to a given quantum state while the main part is
protected against uncontrolled decoherence with the accuracy proportional to
the period. The number of different error Hamiltonians
should not exceed the dimension of the ancilla Hilbert space. Initially
disentangled, the two parts are driven to a highly entangled state
by means of the unitary non-holonomic control. This new state belongs to a
code space such that any error can only move the code vectors in a
direction perpendicular to this space. The inverse unitary
transformation brings the system to a state that differs from the
initial state only in the ancilla part. After resetting the ancilla, the
system returns back to the original state. We propose new
algorithms that give the corresponding code spaces and parameters required
for non-holonomic control. These algorithms have complexity proportional to
the dimension $N$ of the entire Hilbert space. For large systems, the
computational time required to find a code space becomes prohibitive.
However, a coherent control sequence needs only $\log N$ switches to
strongly reduce the decoherence rate and bring the system to a highly
entangled state weakly sensitive to the decoherence.

\end{multicols}


\begin{references}
\bibitem[\dagger]{ad:ak} 
This research was supported by EC QUACS grant.
\bibitem[\ddagger]{ad:ma}This research was supported by NSF grant CCR-0097125.
\bibitem{condition}  Actually, it may differ from zero by a scalar matrix.
We subtract the scalar to make the matrices traceless.
\end{references}
\end{document}